\documentclass[letter]{aa} 

\usepackage{booktabs}
\usepackage{graphicx}
\usepackage{txfonts}
\usepackage{hyperref}

\newcommand{\hi}{\ion{H}{i}}
\newcommand{\mgii}{\ion{Mg}{ii}}

\newcommand{\cluster}{SGAS~J1226+2152}

\newcommand{\puc}{Instituto de Astrof\'\i sica, Pontificia Universidad Cat\'olica de Chile, Casilla 306, Santiago, Chile\label{puc}}
\newcommand{\eso}{European Southern Observatory, Alonso de C\'ordova 3107, Vitacura, Casilla 19001, Santiago, Chile\label{eso}}
\newcommand{\pucv}{Instituto de F\'isica, Pontificia Universidad Cat\'olica de Valpara\'iso, Casilla 4059, Valpara\'iso, Chile\label{pucv}}
\newcommand{\uchile}{Departamento de Astronom\'ia, Universidad de Chile, Casilla 36-D, Santiago, Chile\label{uchile}}
\newcommand{\udp}{Instituto de Estudios Astrof\'isicos, Facultad de Ingenier\'ia y Ciencias, Universidad Diego Portales, Av. Ej\'ercito Libertador 441, Santiago, Chile\label{udp}}

\defcitealias{2021MNRAS.507..663T}{T21}
\newcommand{\tejos}[1]{\citetalias{2021MNRAS.507..663T}}

\begin{document}

\title{Recovering the effective impact parameters in integral-field absorption-line tomography}
\titlerunning{Effective impact parameters in absorption-line tomography}
\authorrunning{J. A. Hern\'andez-Guajardo et al.}

\author{
J. A.~Hern\'andez-Guajardo\inst{\ref{eso},\ref{puc}}
\corrauth{joaquin.hernandezg@uc.cl}
    \and
L. F.~Barrientos\inst{\ref{puc}}
    \and
C.~Ledoux\inst{\ref{eso}}
    \and
N.~Tejos\inst{\ref{pucv}}
    \and
E. J.~Johnston\inst{\ref{udp}}
    \and
P.~Anshul\inst{\ref{uchile}}}

\institute{\eso \and \puc \and \pucv \and \udp \and \uchile}

\date{Received 22 June 2026 / Accepted 5 August 2026}

\abstract
{
Integral-field-unit (IFU) spectroscopy of extended background sources, such as gravitational arcs,
delivers many closely spaced absorption sightlines through the circumgalactic medium (CGM) 
of foreground galaxies. However, the source extent and smearing through the point spread function (PSF) cause
each spaxel to integrate light from a region larger than itself. The usually adopted 
geometric spaxel centre is therefore not representative of where the absorbed flux originates. 
Informed by high-resolution imaging, we introduce a flux-contribution formalism that reconstructs the 
spatial origin of the flux in each spaxel by defining flux-weighted coordinates and 
impact parameters. Using a data-driven simulation built on a real gravitational-arc configuration, 
we show that spaxel centres are displaced from the flux-weighted positions by 
$|\Delta\mathbf{x}|\propto\theta_\text{PSF}\,(1-f)$, with $f$ the spaxel–source overlap fraction,
and that the resulting impact-parameter bias scales with the lensing magnification as $\mu^{-1/2}$. 
These offsets are largest for off-arc spaxels and flatten the radial column-density profiles. They also inject a coherent 
geometry-driven scatter that mimics intrinsic CGM structure and that is removed neither by a tighter 
spaxel selection nor by a finer spatial binning. For identical uncertainties, 
the spaxel-centre coordinates leave residuals of about twice the noise level 
($\chi^2_\nu\!\sim\!4$), whereas the flux-weighted coordinates recover the input profile 
to within the uncertainties ($\chi^2_\nu\!\sim\!1$) and remain robust to a PSF misspecification of $\pm$10\%. The method we propose provides a better-motivated and readily adoptable framework for 
tomographic studies, whether as the primary analysis or as a robustness check on spaxel-centre results.
}


\maketitle
\nolinenumbers
    
\section{Introduction}

Absorption-line spectroscopy against bright background sources has been the primary tool for 
characterising the circumgalactic medium \citep[CGM; see e.g.][]{2017ARA&A..55..389T}. 
In the traditional quasar absorption-line (QAL) approach \citep[e.g.][]{2025qal1.book.....C}, 
each sightline probes intervening gas at a well-defined impact parameter from the foreground absorbing galaxies.
Gravitational-arc tomography \citep{2018Natur.554..493L, 2020MNRAS.491.4442L,2021ApJ...914...92M, 2021MNRAS.507..663T,2022Natur.606...59B,
2025ApJ...986..190S, 2026A&A...708A.376H}
extends this framework by exploiting gravitational arcs as background sources, providing multiple 
sightlines through individual haloes via integral-field-unit (IFU) spectroscopy.

The use of extended background sources, whether a gravitational arc or an 
unlensed galaxy, breaks the pencil-beam assumption from QALs. 
The flux collected in a given resolution element is a superposition of light from across 
the background source, so that its spectrum reflects a flux-weighted average over many sightlines.
In arc tomography, the standard practice is to bin IFU data to a spaxel size similar to
the point spread function (PSF) full width at half maximum (FWHM) and assign absorption to the 
geometric spaxel centre. 
This approach aims to maximise the signal-to-noise ratio (S/N) and the number of spatially independent sightlines. 
However, the spaxel-centre approach assumes that the spaxels are uniformly illuminated.
PSF smearing is known to bias spatially resolved emission-line
gradients \citep[e.g.][]{2013ApJ...767..106Y, 2017MNRAS.468.2140C} and kinematics,
motivating forward-modelling tools that recover intrinsic
quantities \citep[e.g.][]{2015AJ....150...92B}. The impact of PSF smearing
on absorption-line tomography against extended background sources has not been quantified to our knowledge, however.

In this Letter, we introduce a flux-contribution formalism that accounts for PSF smearing and 
the spatial sampling to recover the effective spatial scales probed by each IFU spaxel
in absorption-line tomography. We illustrate the method with 
a data-driven simulation and show that the spaxel-centre approximation leads to measurable 
and systematic biases in the impact parameters, which translate into additional scatter, 
and biases in the slopes and normalisations of CGM radial profiles. 
Throughout, we adopt a flat $\Lambda$CDM cosmology with
$H_0 = 70\ \rm{km\,s^{-1}\,Mpc^{-1}}$ and $\Omega_m = 0.3$.

\section{Flux-contribution formalism}
\label{sec:formalism}

\begin{figure*}
\centering
\includegraphics[width=0.75\linewidth]{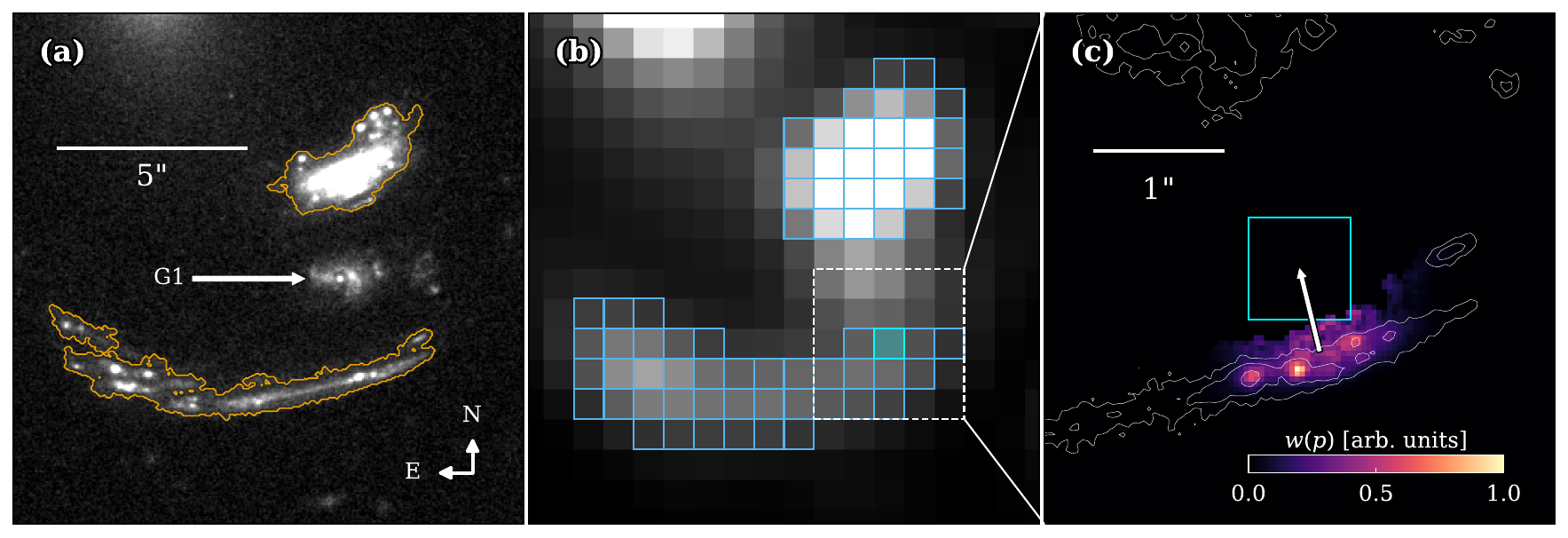}
\caption{
\textit{(a):} HST/F606W image of \cluster{}, with yellow contours delineating gravitational-arc 
regions from a segmentation image.
\textit{(b):} HST/F606W binned and convolved to the PSF FWHM and spaxel size of the binned MUSE observations 
from \tejos{}. The light blue squares mark the spaxels we selected for the analysis; the highlighted cyan square 
indicates an example binned spaxel.
\textit{(c):} Flux-contribution weights $w_a$ for the example spaxel and its corresponding displacement vector 
$\Delta \mathbf{x}$ (white arrow). The white lines show representative HST flux contours. 
}
\label{fig:hst-image-impact-parameter-map}
\end{figure*}

Let $H$ and $L$ be a high- and a low-resolution image, respectively, probing the same 
underlying surface-brightness distribution. The pixel scale of $L$ is strictly larger than that of $H$. 
Let $p$ denote a pixel in $H$ and $a$ a pixel in $L$, with $L(a)$ the flux measured 
in pixel $a$ (analogously for $p$ and $H(p)$).  
If $H$ and $L$ share the same PSF, then each pixel $a$ samples a geometric footprint $M_a$ in 
the grid of $H$, and $L(a) = \sum_p H(p) \, M_a(p)\,$. 
When the PSF of $L$ is broader than that of $H$, however, $L(a)$ also receives contributions from pixels 
outside $M_a$, whose light is redistributed into $a$ by the larger PSF. 
We accounted for this effect by defining the sensitivity map 
$S_a(p)$ (see Appendix~\ref{app:appendix-sensitivity-map}), so that
\begin{equation}
L(a) = \sum_p H(p) \, S_a(p)\,.
\label{eq:conv_pixel_sum}
\end{equation}
To quantify the flux that each pixel $p$ contributes to pixel $a$, we further defined the
flux-contribution map $W_a$ and the normalised flux-contribution weights $w_a$ as 
\begin{equation}
W_a(p) = H(p)\,S_a(p), \qquad w_a(p) = \frac{W_a(p)}{\sum_{p'} W_a(p')}.
\label{eq:flux-contribution-map}
\end{equation}
Since $a$ integrates flux from an extended
region, determined by $w_a$, it does not probe a
pencil-beam sightline. We therefore defined flux-weighted coordinates 
that quantified the effective spatial scales probed by each pixel $a$ for a given 
background-source surface-brightness distribution, instrumental PSF, and spatial sampling. 
We defined the flux-weighted centroid $\bar{\mathbf{x}}_a$ and
its scalar dispersion $\sigma_{x,a}$ as
\begin{equation}
\bar{\mathbf{x}}_a = \sum_p w_a(p)\, \mathbf{x}(p), \qquad \sigma_{x,a}^2 = \sum_p w_a(p)\, | \mathbf{x}(p) - \bar{\mathbf{x}}_a |^2\,,
\label{eq:flux_weighted_coordinate}
\end{equation}
where $\mathbf{x}(p)$
denotes the position vector of pixel $p$. This framework generalises naturally
to impact parameters (Appendix~\ref{app:impact-parameters}) and other spatial coordinates (azimuthal 
angles, effective areas, etc.), and provides a flexible tool for quantifying the physical scales probed 
by IFU observations when 
informed by imaging with a higher spatial resolution. 
Moreover, the overlap of flux-contribution maps between pairs of low-resolution pixels encodes
their pairwise correlations, whose full
treatment we defer to future work.

\section{Data-driven simulation}
\label{sec:simulation}

We applied the formalism to the geometry of the arc-tomography system 
of \citet{2021MNRAS.507..663T}, hereafter \tejos{}: 
\cluster{} \citep{2010ApJ...723L..73K, 2018AJ....155..104R}, a gravitational 
arc at $z\!\approx\!2.9233$ lensed by a cluster at $z\!\approx\!0.43$, whose
spectrum probes \mgii{} absorption from the CGM of a galaxy (G1) 
at $z\!\approx\!0.77$ in Very Large
Telescope (VLT)/Multi Unit Spectroscopic Explorer (MUSE)
\citep{2010SPIE.7735E..08B} data. \tejos{} 
binned the native MUSE spaxels to $0\farcs8$, comparable to their PSF FWHM
($\theta_\text{PSF}\!\sim\!0\farcs7$), and measured absorption strengths and 
kinematics versus de-lensed impact parameter to G1. We adopted their observational
configuration (arc light distribution, PSF, and $0\farcs8$ spaxels) as the basis 
for a data-driven simulation.

We retrieved publicly available \textit{Hubble} Space Telescope (HST) F606W imaging from the
Mikulski Archive for Space Telescopes\footnote{\url{https://archive.stsci.edu/}},
whose bandpass covers the
observed wavelength of \mgii{} absorption at $z\!\sim\!0.77$ ($\approx\!
5000$~\AA). Hereafter, we refer to this high-resolution image as $H$; 
it has $\theta^\text{H}_\text{PSF}\approx\!0\farcs1$ and a
pixel scale of $0\farcs04$~pix$^{-1}$, and is shown in panel~a of 
Fig.~\ref{fig:hst-image-impact-parameter-map}. To mimic the background arc
continuum around the \mgii{} detections reported by \tejos{}
as it would appear in the binned MUSE data\footnote{Assuming negligible
colour gradients and detector effects.},
we convolved $H$ with a Gaussian kernel with an FWHM 
$0\farcs693$ to match $\theta_\text{PSF}$, and then
applied a $0\farcs8$ binning
to construct the low-resolution image $L$. 
For convenience, we hereafter
refer to the low-resolution pixels in $L$ as spaxels.

We then constructed a segmentation mask defining the regions in $H$ that correspond to
the background gravitational arc. Its contours are shown
in panel~a of Fig.~\ref{fig:hst-image-impact-parameter-map}. Next, we selected 
spaxels with $>50\%$ of the integrated flux from the arc and an integrated 
$\text{S/N}\!>\!2$ over the resulting
background sky noise. This threshold was deliberately relaxed. It reproduces
the \tejos{} spaxels and adds fainter spaxels that are dominated by PSF-smeared flux, 
which retain measurable arc flux but are expected to be most strongly biased. 
The $L$ image and our selected spaxels are shown in Fig.~\ref{fig:hst-image-impact-parameter-map}.

Finally, we constructed the sensitivity maps $S_a$ (see Appendix~\ref{app:appendix-sensitivity-map} and
Fig.~\ref{fig:mask-sensitivity-map-psf}) 
and computed the corresponding 
flux-contribution weights $w_a$ via
Eq.~\eqref{eq:flux-contribution-map} for each selected spaxel $a$. 
For each spaxel, we computed its effective
image-plane coordinates $\bar{\mathbf{x}}_a$ using 
Eq.~\eqref{eq:flux_weighted_coordinate}, and we calculated the 
overlap fraction $f$, defined as the fraction of each spaxel's geometric 
footprint overlapping with the arc mask in the unconvolved HST 
segmentation mask (Fig.~\ref{fig:hst-image-impact-parameter-map}, panel~a). 
The geometry and observational setup are thus entirely
empirical, allowing us to compare spaxel-centre and flux-weighted centroids. 
Panel~c of Fig.~\ref{fig:hst-image-impact-parameter-map}
shows an example $w_a$, illustrating that for some spaxels, particularly
those lying just outside the arc continuum, most of the flux 
originates from beyond their own footprint. 

\section{Results and discussion}
\label{sec:results-discussion}
\subsection{Flux-weighted centroid displacements}

\begin{figure}
\centering
\includegraphics[width=\linewidth]{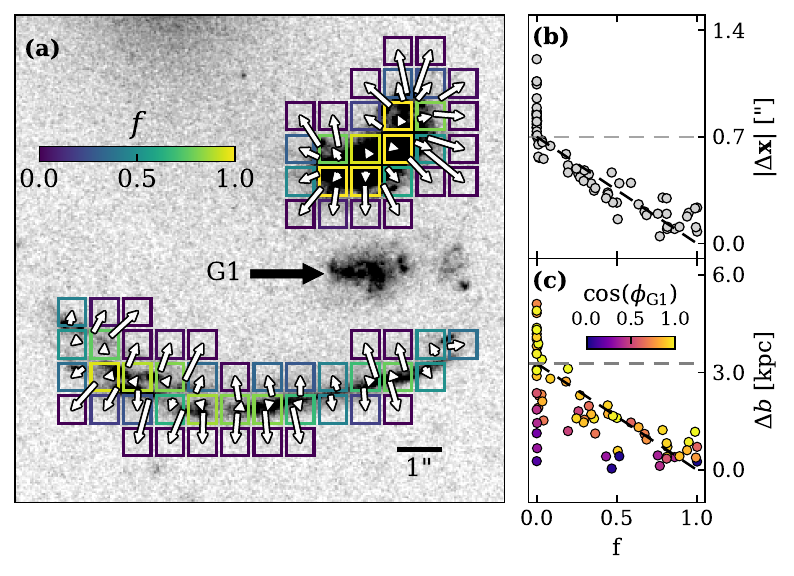}
\caption{
\textit{(a):} 
Native-resolution HST/F606W image with spaxels (open squares, coloured by the overlap fraction $f$) 
and their displacement vectors $\Delta \mathbf{x}$ (white arrows, drawn to scale). 
The black arrow points towards the G1 galaxy.
\textit{(b):}
Magnitude of the displacement vectors, $|\Delta \mathbf{x}|$, vs. $f$. 
The dashed grey line indicates $\theta_\text{PSF}\!=\!0\farcs7$, 
and the dashed black line shows the relation $|\Delta\mathbf{x}|\!=\!\theta_\text{PSF}\times(1-f)$. 
\textit{(c):}
Impact-parameter bias $|\Delta b|$ vs. $f$,
coloured by $\cos{(\phi_\text{G1})}$, where $\phi_\text{G1}$ is the relative 
angle between $\Delta \mathbf{x}$ and the direction to the centre of G1.
The dashed black line indicates $|\Delta b|_\text{max}$ from
Eq.~\eqref{eq:scaling} for $f=0$--$1$,  and the dashed grey line indicates the respective value for $f=0$.
}
\label{fig:displacement-vectors}
\end{figure}

The displacement vectors, $\Delta \mathbf{x}=\mathbf{x}^{\mathrm{centre}}-\bar{\mathbf{x}}$, describe
how the spaxel centres are displaced from their flux-weighted centroids (panel~a of
Fig.~\ref{fig:displacement-vectors}). For on-arc spaxels ($f > 0$),
$|\Delta\mathbf{x}|$ decreases linearly with increasing $f$. 
As $f\rightarrow 1$, $|\Delta\mathbf{x}|\rightarrow0$, and as $f\rightarrow 0$,
$|\Delta\mathbf{x}|\rightarrow\theta_\text{PSF}$. A simple linear relation 
$|\Delta\mathbf{x}|=\theta_\text{PSF}\times(1-f)$ captures the observed trend,
as shown in panel~b of Fig.~\ref{fig:displacement-vectors}. 
Off-arc spaxels ($f = 0$), in contrast, show
$|\Delta\mathbf{x}|\!\gtrsim\!\theta_\text{PSF}$ as their integrated flux 
is entirely given by PSF-smeared light from the arc. 

To translate these displacements into impact-parameter biases, we constructed an
impact-parameter map relative to G1, forward-modelled to the image plane (in the
same pixel grid as $H$) using the strong-lensing model from \tejos{}
(see Appendix~\ref{app:impact-parameters}), and applied
Eq.~\eqref{eq:effective_impact_parameter} to obtain the flux-weighted impact
parameters $\bar{b}_a$ and the dispersion $\sigma_{b,a}$ for each spaxel $a$. The
spaxel-centre value $b^{\mathrm{centre}}_a$ was evaluated at the geometric centre
of each spaxel (following \tejos{}), with uncertainties estimated from the range
of impact parameters spanned by the spaxel footprint.

The impact-parameter bias $\Delta b = \bar{b}-b^{\mathrm{centre}}$ is set by $|\Delta \mathbf{x}|$ but 
also by the orientation of $\Delta\mathbf{x}$ relative to G1, defined as
$\phi_\text{G1}$: at fixed $f$, $|\Delta b|$ is largest when $\Delta \mathbf{x}_a$
points towards or away from G1, and vanishes when it is perpendicular (panel~c of
Fig.~\ref{fig:displacement-vectors}). Since the local lensing magnification $\mu$
over the arc (at the sightline position) reduces the physical scale of an angular
separation by $\sqrt{\mu}$ (approximating $\mu$ as isotropic), the expected
maximum bias at a given $f$ is
\begin{equation}
|\Delta b|_\text{max} \approx |\Delta \mathbf{x}| \, s(z) \, \mu^{-1/2}
    = \theta_\text{PSF} \, (1-f) \, s(z) \, \mu^{-1/2},
    \label{eq:scaling}
\end{equation}
where $s(z)$ is the physical angular scale at redshift $z$ (the relations also
hold in the unlensed limit, i.e. when $\mu=1$). For the \tejos{} system ($\mu\!\sim\!2.5$ and
$s\!\approx\!7.4$~kpc/\arcsec\ 
at $z\!=\!z_\text{G1}$), Eq.~\eqref{eq:scaling} yields $|\Delta b|_\text{max}\!\approx\!0$--$3.3$~kpc 
for on-arc spaxels, and for $|\Delta\mathbf{x}|\lesssim1.5\,\theta_\text{PSF}$,
$|\Delta b|\lesssim5$~kpc for off-arc ones. The off-arc spaxels therefore carry
the strongest biases, precisely the regime where the spaxel-centre approximation is
least justified.

\subsection{Impact on radial profiles}
\label{sec:impact-radial-profiles}

\begin{figure}
\centering
\includegraphics[width=\linewidth]{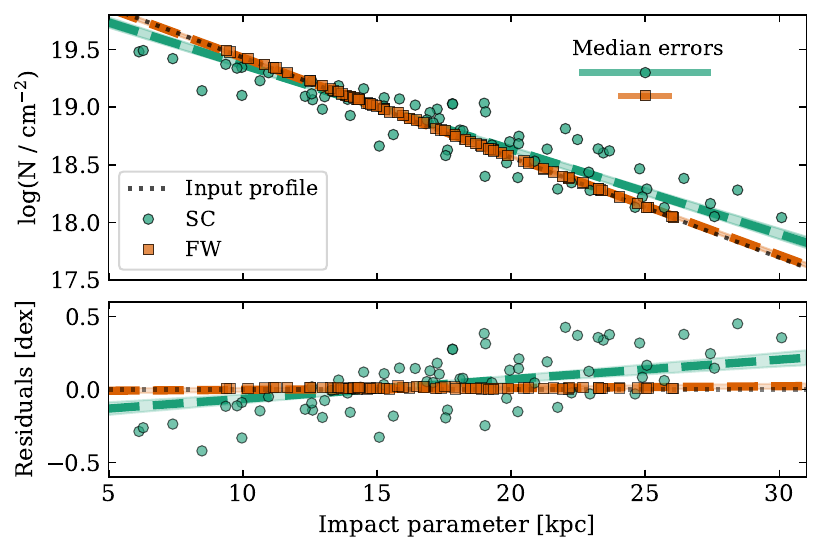}
\caption{
\textit{Top:} 
Simulated column density vs. impact
parameter. The green circles and orange squares represent the SC and FW datasets, respectively.
Impact-parameter error bars are omitted for clarity; the median
errors are shown in the top right corner. The dashed green and orange lines
show the best-fit exponential models, and the dotted black line
shows the input model.
\textit{Bottom:} Column-density residuals relative to
the input model. The colour scheme and line styles
are the same as above.
}
\label{fig:radial-profile-results}
\end{figure}

\begin{table}[t]
\footnotesize
\centering
\caption{Recovery of the input parameters ($b_e\!=\!5.0$~kpc; $\log N_0/\mathrm{cm}^{-2}\!=\!20.3$) from noiseless column densities using the SC and FW datasets.}
\label{tab:sp_fw:noiseless}
\begin{tabular}{lccc}
\toprule
Method & $b_e$ & $\log N_0/\mathrm{cm^{-2}}$ & $\langle\lvert\Delta\log N\rvert\rangle$  \\
& (kpc) &  & (dex)  \\
\midrule
SC & $5.7 \pm 0.3$ & $20.14 \pm 0.05$ & $0.13$  \\
FW & $4.99 \pm 0.07$ & $20.31 \pm 0.02$ & $0.01$ \\
\bottomrule
\end{tabular}
\end{table}
\begin{table}[t]
\footnotesize
\centering
\caption{Same as Table~\ref{tab:sp_fw:noiseless}, but with mock uncertainties 
in $\log \bar{N}_a$. }
\label{tab:sp_fw:noisy}
\begin{tabular}{lcccc}
\toprule
Method & $b_e$ & $\log N_0/\mathrm{cm^{-2}}$ & $\langle\lvert\Delta\log N\rvert\rangle$ & $\chi^2_\nu$ \\
& (kpc) &  & (dex) & \\
\midrule
SC & $5.9 \pm 0.6$ & $20.1 \pm 0.3$ & $0.17$ & $4.0 \pm 0.5$ \\
FW, $0.9\theta_{\rm PSF}$ & $5.0 \pm 0.5$ & $20.3 \pm 0.4$ & $0.11$ & $1.1 \pm 0.2$ \\
FW, $1.0\theta_{\rm PSF}$ & $5.0 \pm 0.5$ & $20.3 \pm 0.3$ & $0.11$ & $1.0 \pm 0.2$ \\
FW, $1.1\theta_{\rm PSF}$ & $5.0 \pm 0.5$ & $20.3 \pm 0.3$ & $0.11$ & $1.0 \pm 0.2$ \\
\bottomrule
\end{tabular}
\tablefoot{For FW, we varied the assumed PSF FWHM by 
$\pm 10\%$ about the fiducial value used to construct $L$. $\chi^2_\nu$ quantifies the quality of the fit.}
\end{table}

To compare the effect of adopting $\bar{b}$ or $b^\text{centre}$ on the radial profiles, we considered
a hypothetical case in which G1 was surrounded by an extended \hi{} 
halo with a unity covering fraction and an exponential column-density profile $N(b)\!=\!N_0e^{-b/b_e}$. We 
adopted $N_0\!=\!10^{20.3}$~cm$^{-2}$ and $b_e\!=\!5$~kpc, based on the 
canonical damped Lyman-$\alpha$ (DLA)
threshold (\citealt{2005ARA&A..43..861W}) and the confinement 
of the DLA-strength \hi{} to impact parameters of a few kiloparsecs \citep{2017MNRAS.469.2959K}. We then
computed the spaxel-wise column density $\bar{N}_a$ by integrating a 2D column-density
map, weighted by $w_a$, as detailed in
Appendix~\ref{app:radial-profile-simulation}. This treats $N(b)$ as optically
thin, an idealisation we adopted to isolate the purely geometric biases. Saturated
lines such as \mgii{} violate this assumption, and an appropriate modelling of
the optical depth is then needed
(Appendix~\ref{app:flux-weighting-absorption-lines}).

Figure~\ref{fig:radial-profile-results} shows $\bar{N}$ versus $\bar{b}$ and
$b^\text{centre}$. We performed orthogonal distance
regression \citep[ODR;][]{doi:10.1137/0908085} fits to both datasets (hereafter
FW and SC, referring to the $\bar{N}$--$\bar{b}$ and $\bar{N}$--$b^\text{centre}$
datasets, respectively), accounting for errors in $b$ and zero error in $\bar{N}$
(the model is noiseless). Table~\ref{tab:sp_fw:noiseless} summarises our results. 
The FW best-fit
parameters recover the input values, whereas the SC fit overestimates the scale
length by $15\%$ while
underestimating $\log N_0$ by $0.16$~dex; that is, the profile is flattened and renormalised.
Since the model is noiseless, this bias cannot come from $\bar{N}$: it must
originate entirely from the misassignment of $b^\text{centre}$ as the impact
parameter. The
resulting residuals reach $\sim0.5$~dex, biased towards lower $N$ at small $b^\text{centre}$ and
higher $N$ at large $b^\text{centre}$. This pattern follows directly from the orientation of
the $\Delta \mathbf{x}$ vectors shown in Fig.~\ref{fig:displacement-vectors}:
flux from the arc edges facing G1 is PSF-smeared towards G1 (hence to smaller
$b^\text{centre}$), while arc edges facing away from G1 are smeared in the
opposite direction.

The same geometric bias persists under realistic observational conditions. We injected
mock uncertainties in $\bar{N}$ (Appendix~\ref{app:radial-profile-simulation})
and repeated the fitting procedure, additionally varying the $\theta_\text{PSF}$
used to reconstruct the $w_a$ maps by $\pm10\%$ to mimic systematics in the PSF
(Table~\ref{tab:sp_fw:noisy}). The FW fits recover the input parameters within
$1\!\sigma$ with a mean absolute residual of $0.11$~dex and
$\chi^2_\nu\approx1$, regardless of $\theta_\text{PSF}$. The SC fit yields
parameters consistent with the input model within 
$1.7\!\sigma$ in $b_e$ and $1\!\sigma$ in $\log N_0$,
albeit with a significantly larger mean absolute residual of $\approx\!0.17$~dex
and $\chi^2_\nu\approx4$ (a roughly four-fold increase), indicating excess
scatter beyond the injected noise. Because the column-density uncertainties are
identical, this excess scatter is driven by the impact-parameter misassignment,
and it is therefore geometric in origin. 

A tighter spaxel selection did not remove these geometric biases. Raising 
the S/N
cut leaves or even worsens the bias in SC, while raising $f$ recovers
the profile
shape, but leaves $\chi^2_\nu \approx 4$--$6$ 
(Appendix~\ref{app:robustness}). The FW fit is
insensitive to both, with $\chi^2_\nu \approx 1$ throughout. Finer 
spaxel scales did not improve the recovery of the radial profiles either.
Even at the native $0\farcs2$ sampling, the spaxel-centre fit still 
overestimates $b_e$ (by $14\%$) and underestimates $\log N_0$ (by 
$\sim0.17$~dex) because when the spaxel is smaller than the PSF, the 
sampled region saturates at the PSF width, and finer spaxel scales yield
no effective gain in resolution (Appendix~\ref{app:binning-test}).

The accuracy of $\bar{b}$ depends on how well the high-resolution 
imaging represents the IFU continuum at the wavelength of interest 
and on the adopted PSF model. The former is readily checked from 
colour gradients and from the absence of strong emission lines in 
the IFU data within the imaging passband; the latter is subdominant,
since a $\pm10\%$ error in $\theta_\text{PSF}$ leaves the fits unchanged.
In addition, lens models carry their own systematics, which should be
appropriately propagated into the impact parameters. However, these 
systematics cancel in our comparison because $\bar{b}$ and $b^\text{centre}$
derive from the same model. In the unlensed regime, no systematics enter beyond
the adopted cosmology.

\section{Conclusions}
\label{sec:conclusions}

We have presented a formalism that accounts for PSF smearing and spatial
sampling when studying absorption lines against extended background 
sources.
Using a data-driven simulation built on the real \tejos{} system, we found
that 
(i) in the image plane, spaxel-centre coordinates are displaced by
$|\Delta\mathbf{x}|=\theta_\text{PSF}\,(1-f)$ for on-arc spaxels and
$|\Delta\mathbf{x}|\gtrsim\theta_\text{PSF}$ for the off-arc ones, 
translating into impact-parameter biases
$|\Delta b|_\text{max}\approx\theta_\text{PSF}\,(1-f)\,s(z)\,\mu^{-1/2}$
for on-arc spaxels and up to
$|\Delta b|\approx1.5\,\theta_\text{PSF}\,s(z)\,\mu^{-1/2}$ for off-arc
spaxels (vanishing when $\Delta\mathbf{x}$ is perpendicular to the
G1 direction),
for local lensing magnification $\mu$, with $\mu=1$ in the no-lensing 
regime;
(ii) the radial profiles inferred by adopting spaxel centres are biased
in shape ($b_e$ overestimated by $15\%$ here and $\log N_0$ underestimated
by $0.16$~dex) and carry a correlated 
geometry-driven scatter (low $N$ at small
$b$ and vice versa) that mimics intrinsic CGM structure; 
(iii) the
flux-weighted coordinates recover the input profile within $1\sigma$ with
$\chi^2_\nu\approx1$, consistent with the injected noise, and are
robust to $\pm10\%$ PSF misspecification, whereas the spaxel-centre fit
reaches $\chi^2_\nu\approx4$ on identical uncertainties, that is, a scatter 
excess
that is purely geometry-driven; 
and (iv) cuts in the spaxel--arc overlap fraction improve the recovery of the
input radial profile shape, but neither tighter S/N cuts nor
overlap-fraction cuts remove the artificial scatter introduced
by the spaxel-centre approximation. 

The framework we presented is not restricted to the
optically thin regime. The same $w_a$ maps apply to saturated transitions
when the optical depth is modelled explicitly
(Appendix~\ref{app:flux-weighting-absorption-lines}). This formalism is
inexpensive to apply when high-resolution imaging is available, and we 
encourage tomographic studies to adopt it
either as the primary analysis or as a robustness check of the spaxel-centre
results. 

\section{Data availability}
The Python code implementing the flux-contribution formalism and
the simulation presented in this Letter is publicly available 
on GitHub at \url{https://github.com/joaquinhernandezg/flux-weighted-impact-parameters}.

\begin{acknowledgements}
We thank the anonymous referee for their constructive comments, which improved
the quality of this work. J.A.H. acknowledges support from the European Southern Observatory (ESO)
through the ESO Studentship programme,
and support from Beca ANID Doctorado Nacional Folio 21230522.
J.A.H. and L.F.B. acknowledge support from ANID BASAL project FB210003,
FONDECYT project 1230231,
and CASSACA project CCJRF1906.
N.T. acknowledges support by FONDECYT grant 1231187.
E.J.J. acknowledges support by FONDECYT Regular grant number 1262304 and 
ANID CATA-BASAL project FB210003. P.A. acknowledges support from
ANID-Subdirecci\'on de Capital Humano/Doctorado Nacional/2022-21222110.
Based on observations made with the NASA/ESA \textit{Hubble} Space Telescope.  
\end{acknowledgements}

\bibliographystyle{aa}
\bibliography{references.bib}

\begin{appendix}
\nolinenumbers

\section{Sensitivity map}
\label{app:appendix-sensitivity-map}

In this appendix, we work with explicit pixel coordinates, which relate to the 
notation of Sect.~\ref{sec:formalism} as follows: a low-resolution pixel $a$ 
corresponds to the coordinate pair $(x,y)$ in $L$, and a high-resolution pixel $p$ 
to the pair $(i,j)$ or $(u,v)$ in $H$, so that $M_a = M_{x,y}$ and $S_a(p) = S_{x,y}
(i,j)$. Let $M_{x,y}$ be a binary mask defined in $H$, 
selecting the set of pixels that geometrically overlap with the low-resolution pixel
$(x, y)$ in  $L$. Given a convolution kernel $K$ that matches the PSF of 
$H$ to that of $L$, the measured flux in pixel $(x, y)$ can be written as
\begin{equation}
    L(x,y) = \sum_{i,j} M_{x,y}(i,j)\,\left(H \otimes K\right)(i,j)\,.
\end{equation}
Expanding the convolution,
\begin{equation}
    \left(H \otimes K\right)(i,j) = \sum_{u,v} H(u,v)\,K(i-u, j-v)\,,
\end{equation}
where $(u,v)$ index the pixels of the high-resolution image.
Substituting into the previous expression,
\begin{equation}
    L(x,y) = \sum_{i,j} M_{x,y}(i,j) \sum_{u,v} H(u,v)\,K(i-u, j-v)\,.
\end{equation}
Swapping the order of summation,
\begin{align}
    L(x,y)
    &= \sum_{u,v} H(u,v) \sum_{i,j} M_{x,y}(i,j)\,K(i-u, j-v) \\
    &= \sum_{u,v} H(u,v)\,S_{x,y}(u,v)\,,
\end{align}
where we defined the sensitivity map
\begin{equation}
    S_{x,y}(u,v) = \sum_{i,j} M_{x,y}(i,j)\,K(i-u, j-v)\,.
\end{equation}
This can be written as a convolution with the flipped kernel,
\begin{equation}
    S_{x,y} = M_{x,y} \otimes K^\dagger\,,
\end{equation}
where $K^\dagger(u,v) = K(-u,-v)$. If the PSF kernel is symmetric, $K^\dagger=K$, and therefore
\begin{equation}
    S_{x,y} = M_{x,y} \otimes K\,.
    \label{eq:sensitivity-map}
\end{equation}

\begin{figure}[h!]
\centering
\includegraphics[width=\linewidth]{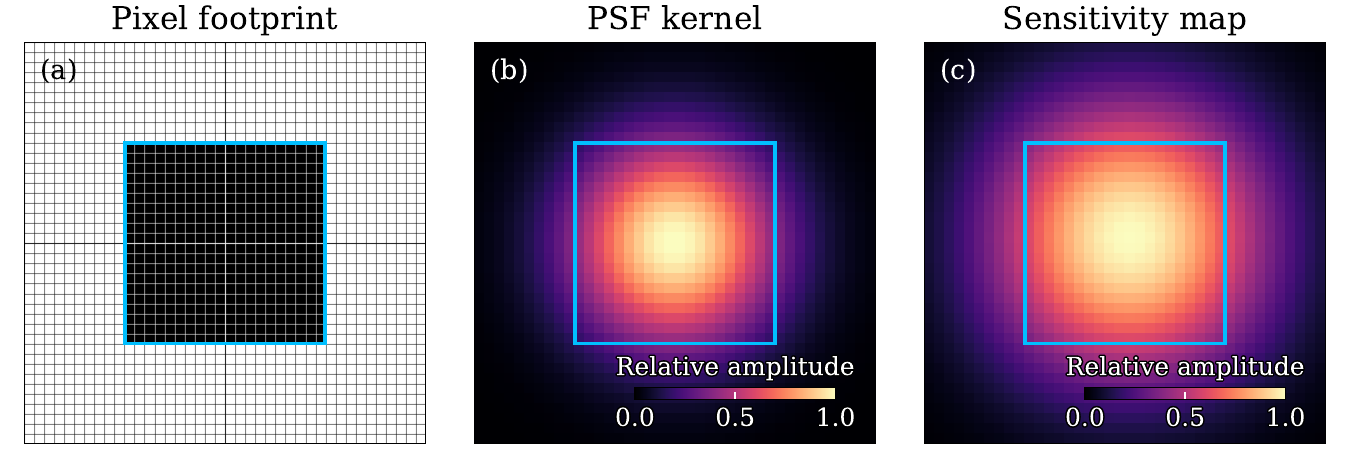}
\caption{Construction of the sensitivity map $S_a$ for a single low-resolution pixel.
\textit{(a):} Geometric mask $M_a$ on the high-resolution grid (black pixels: inside;
white pixels: outside), with the blue square ($0\farcs8$ side) marking the
low-resolution pixel footprint.
\textit{(b):} Gaussian PSF convolution kernel $K$ with $\theta_\text{PSF}=0\farcs7$.
\textit{(c):} Resulting sensitivity map $S_a=M_a\otimes K$
(Eq.~\eqref{eq:sensitivity-map}), whose response extends well beyond the geometric
footprint. The blue square is repeated in all panels, and the colour scales in panels
(b) and (c) are normalised to their peak values.
}
\label{fig:mask-sensitivity-map-psf}
\end{figure}

\section{Impact-parameter map and effective impact parameters}
\label{app:impact-parameters}

\subsection{Source-plane impact parameters}
\label{app:impact-parameters-lensing}

A common product of strong-lensing models is the deflection field $\boldsymbol{\alpha}(\boldsymbol{\theta})$,
which maps a position in the image plane to its corresponding position in the source plane through the lens equation
\begin{equation}
    \boldsymbol{\beta} = \boldsymbol{\theta}- \boldsymbol{\alpha}(\boldsymbol{\theta}; z_s)\,,
    \label{eq:lens-equation}
\end{equation}
where $\boldsymbol{\theta}=(\theta_\alpha, \theta_\delta)$ denotes an angular position in the image plane
and $\boldsymbol{\beta}=(\beta_\alpha, \beta_\delta)$ the corresponding position in the source plane. The deflection field
$\boldsymbol{\alpha}(\boldsymbol{\theta}; z_s)$ is evaluated at the image-plane position $\boldsymbol{\theta}$, and 
appropriately scaled to the source-plane redshift $z_s$.

To compute the
impact parameter with respect to a given galaxy G, we calculated the physical transverse distance between
the de-lensed pixel position $\boldsymbol{\beta}$ and the de-lensed position of the galaxy centre
$\boldsymbol{\beta}_G$ as
\begin{equation}
    b = d_A(z_\mathrm{G})\,\Delta \boldsymbol{\beta}\,,
    \label{eq:transverse-distance}
\end{equation}
where $d_A(z_\mathrm{G})$ is the angular-diameter distance at redshift $z_\mathrm{G}$ 
and $\Delta \boldsymbol{\beta}$ denotes the
angular separation between $\boldsymbol{\beta}$ and $\boldsymbol{\beta}_\mathrm{G}$.

\subsection{Impact-parameter map}
\label{app:impact-parameter-map}

\begin{figure}[htb!]
\centering
\includegraphics[width=0.6\linewidth]{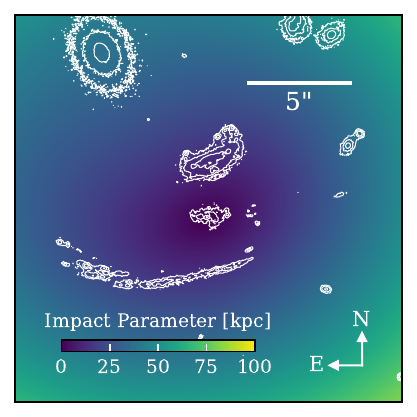}
\caption{
Impact-parameter map for the \tejos{} system computed on the pixel
grid of the HST high-resolution image. Each pixel $p$ was mapped
to the absorber plane at the redshift of G1 ($z\approx 0.77$) using
the strong-lensing deflection field and
Eqs.~\eqref{eq:lens-equation} and \eqref{eq:transverse-distance}.
The colour scale shows the physical transverse distance $b(p)$ to
the de-lensed position of G1. The white contours correspond
to representative flux levels from the
HST/F606W image for spatial reference.
}
\label{fig:impact-parameter-map}
\end{figure}

Using this transformation, each image-plane pixel $p$ can be mapped
to a position $\boldsymbol{\beta}(p)$ in the source plane, yielding
a per-pixel impact parameter $b(p)$ via
Eqs.~\eqref{eq:lens-equation} and \eqref{eq:transverse-distance}.
Unlike the conventional approach of de-lensing the observed image to
reconstruct a source-plane representation, we instead assigned to each
pixel its ray-traced impact parameter with respect to G1, producing
an impact-parameter map defined on the native high-resolution pixel
grid. We computed this map for the \tejos{} system on the
high-resolution HST pixel grid described in
Sect.~\ref{sec:simulation}, measuring impact parameters relative to
the absorbing galaxy G1 at $z\approx 0.77$. The resulting map is
shown in Fig.~\ref{fig:impact-parameter-map}.

\subsection{Effective impact parameters}
\label{app:effective-impact-parameters}

As a concrete application
relevant to absorption-line studies, we considered
the effective impact parameter. From Eq.~\eqref{eq:flux_weighted_coordinate}, 
the effective impact parameter for a low-resolution pixel $a$ and its corresponding
dispersion are
\begin{equation}
\bar{b}_a = \sum_p w_a(p)\, b(p)\,, \qquad \sigma_{b,a}^2 = \sum_p w_a(p)\, \left[ b(p) - \bar{b}_a \right]^2\,
\label{eq:effective_impact_parameter}
,
\end{equation}
where the sum runs over all the high-resolution pixels $p$, $w_a(p)$ are the flux-contribution 
weights for pixel $a$, defined in Eq.~\eqref{eq:flux-contribution-map}, and $b(p)$ is the 
impact parameter of pixel $p$ obtained from Eq.~\eqref{eq:transverse-distance}.

\section{Absorption against extended background sources}
\label{app:flux-weighting-absorption-lines}

Let $\tau(p,\lambda)$ be the optical depth along the sightline through pixel
$p$ at wavelength $\lambda$, and let $H(p)$ be its continuum flux. 
We assumed that the pixel scale is fine enough that spatial variations
in $\tau$ occur over scales larger than a single pixel. 
An extended absorbing gas cloud attenuates each
high-resolution sightline independently, producing a monochromatic absorbed
image
\begin{equation}
  h(p,\lambda)=H(p)\,e^{-\tau(p,\lambda)}.
\end{equation}
This image is redistributed into the low-resolution spaxels by the same
sensitivity map $S_a(p)$ (Appendix~\ref{app:appendix-sensitivity-map}) that relates the continuum 
to the binned image in Eq.~\eqref{eq:conv_pixel_sum}. The monochromatic
flux reaching spaxel $a$ is therefore
\begin{equation}
  F_a(\lambda)=\sum_p S_a(p)\,h(p,\lambda)=\sum_p W_a(p)\,e^{-\tau(p,\lambda)},
\end{equation}
where the weight $W_a(p)=H(p)\,S_a(p)$ is exactly the flux-contribution map of
Eq.~\eqref{eq:flux-contribution-map}. In the absence of absorption ($\tau=0$)
this reduces to the continuum $\sum_p W_a(p)=L(a)$, so normalising by
it gives the measured transmission
\begin{equation}
  T_a(\lambda)=\frac{F_a(\lambda)}{\sum_p W_a(p)}=\sum_p w_a(p)\,e^{-\tau(p,\lambda)}.
  \label{eq:effective-transmission}
\end{equation}
The observed transmission is thus the flux-weighted mean of the per-sightline transmissions, 
carrying the same weights $w_a$ that define the centroids $\bar{\mathbf{x}}$ 
in Eq.~\eqref{eq:flux_weighted_coordinate}.

In the optically thin limit, $e^{-\tau}\approx1-\tau$,  with $\tau$ proportional to the column 
density $N$. Integrating the apparent optical depth $-\ln T_a$ over the absorption line yields 
an effective column density 
\begin{equation}
\bar{N}_a=\sum_p w_a(p)\,N(p).
\label{eq:effective_column_density}
\end{equation}
Therefore, $\bar b_a$ (Eq.~\eqref{eq:effective_impact_parameter}) and $\bar{N}_a$ 
inherit identical continuum-flux weighting.
Saturation breaks the $\tau$--$N$ linearity, and the correct approach 
is to adopt a $\tau(p,\lambda)$ model and use Eq.~\eqref{eq:effective-transmission} instead.

\subsection{The effect of a clumpy absorber}

A clumpy absorbing medium does not affect the formalism itself:
Eq.~\eqref{eq:effective-transmission} follows from the linear superposition of
flux and assumes nothing about the spatial structure of $\tau$. What clumpiness
alters is the interpretation of $\bar{N}_a$.

The relevant comparison is between the clump size $r_{\rm cl}$ and the transverse
scale over which each spaxel averages, $\ell_{\rm eff}$. For our fiducial configuration, the
flux-weighted dispersion of Eq.~\eqref{eq:effective_impact_parameter} has a median
$\sigma_b\approx1$~kpc. We characterised the effective transverse scale over which flux is weighted as $\ell_{\rm eff} \approx 2.4$~kpc, corresponding to the FWHM of this distribution. For reference, $\theta_\text{PSF}=0\farcs7$ 
subtends $5.2$~kpc in the image plane and $3.3$~kpc once de-lensed by $\mu^{-1/2}$. Cool CGM gas traced by \mgii{} is observed to be structured on $\sim\!1$--$10$~kpc 
scales \citep[e.g.][]{2018ApJ...868..142R,2023A&A...680A.112A,
2024MNRAS.528.1895D}, that is, comparable to $\ell_{\rm eff}$. Three regimes 
follow. For $r_{\rm cl}\!\gg\!\ell_{\rm eff}$, the medium is locally
uniform and $\bar{N}_a$ is unbiased. For $r_{\rm cl}\!\ll\!\ell_{\rm eff}$, many clumps
are averaged within each spaxel, and the scatter is correspondingly suppressed.
The intermediate case, $r_{\rm cl}\!\sim\!\ell_{\rm eff}$, is where we expect
the strongest effects. 
Here, only a few clumps contribute to each weighted average, so $\bar{N}_a$
fluctuates at the order-unity level between spaxels and should be read as a
transmission-weighted mean. For saturated transitions, $\bar{N}$ is
additionally biased towards lower values, because low-$\tau$ gaps dominate the
emergent flux; partial covering, however, remains constrainable from doublet
ratios and residual flux.

The practical consequence concerns the scatter. Clumpiness contributes a
stochastic term to the $N$--$b$ relation, on top of the coherent,
geometry-driven scatter quantified in this Letter. The two are separable in
principle, since only the geometric term is orientation-dependent, correlating
with $f$ and with the angle of $\Delta\mathbf{x}$ relative to the absorbing
galaxy (panel c of Fig.~\ref{fig:displacement-vectors}). The residual scatter is then
directly informative about $r_{\rm cl}$ and the covering fraction. A
quantitative treatment, however, requires a specific
$\tau(p,\lambda)$ model for these clumps,
which lies beyond the scope of this Letter.

\section{Column-density-profile simulation}
\label{app:radial-profile-simulation}

\subsection{Column-density model}
\label{app:column-density-model}

\begin{figure}[]
\centering
\includegraphics[width=0.6\linewidth]{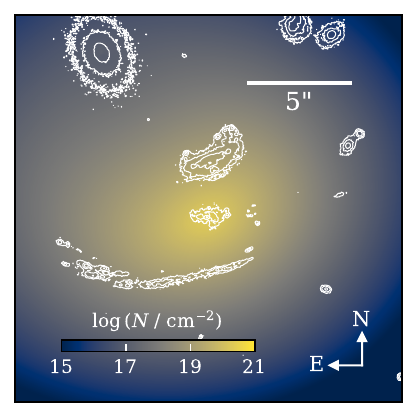}
\caption{Simulated image-plane column-density model for the absorbing galaxy
G1. The white contours correspond to representative flux levels from the
HST/F606W image for spatial reference.}
\label{fig:column-density-map}
\end{figure}

We adopted the exponential column-density profile
$N(b)=N_0\,e^{-b/b_e}$ of Sect.~\ref{sec:impact-radial-profiles}, evaluated on
the impact-parameter map described in
Appendix~\ref{app:impact-parameter-map}, to produce an image-plane column-density map, 
shown in Fig.~\ref{fig:column-density-map}. It is a
deliberately simple, monotonically declining toy model, with no azimuthal
structure and no patchiness. We further assumed that $N(b)$ varies negligibly
across a high-resolution pixel, and that the absorber is a
geometrically thin screen, so that each high-resolution sightline is fully characterised 
by a single impact parameter and has no line-of-sight kinematics.
These assumptions ensure that any scatter or bias in the recovered profiles
arises solely from the geometric effect quantified in this Letter.

\subsection{Mock column densities and uncertainties}
\label{app:mock-noise}

We simulated the measured column density $\bar{N}_a$ in each spaxel using
Eq.~\eqref{eq:effective_column_density}. To mimic observational uncertainties,
we added a log-normal scatter,
$\log\bar{N}^\mathrm{obs}_a=\log\bar{N}_a+\mathcal{N}(0,\sigma_a)$, scaling
with the spaxel flux as
$\sigma_a=\sigma_\mathrm{ref}\sqrt{F_\mathrm{ref}/F_a}$, so that fainter
regions are assigned lower S/N. The scatter was anchored to a reference flux
$F_\mathrm{ref}$ matching the brightest selected spaxel, with reference
dispersion $\sigma_\mathrm{ref}\approx0.02$~dex, giving a mean per-spaxel
uncertainty across the selected spaxels of $\bar{\sigma}\approx0.2$~dex.

\section{Sensitivity to PSF misspecification, spaxel selection, and spaxel size}
\label{app:robustness}

\subsection{Fitting and parameter-uncertainty estimation}
\label{app:fit-procedure}

We fitted the exponential column-density profile
$\log N(b) = \log N_0 - b/(b_e\ln 10)$, with $k = 2$ free parameters (the
central column density $N_0$ and the scale length $b_e$).  We report two
experiments. The first (Table~\ref{tab:sp_fw:noiseless}) is an ODR fit to the noiseless mock column densities, to measure how
faithfully the input model is recovered in the best-case scenario of no observational errors.
We estimated the parameter uncertainties by sampling the error in the impact parameters in 
$3\,000$ realisations.
For the second experiment (Table~\ref{tab:sp_fw:noisy}), we repeated the ODR fit, 
this time sampling both the impact parameters and column densities within their mock uncertainties, 
in $3\,000$ realisations. 
Additionally, we varied the assumed PSF size used to reconstruct the FW impact parameters
(Appendix~\ref{app:psf-test}).

For the second experiment, we tested whether the residual scatter 
is consistent with the simulated
measurement errors. For each realisation we evaluated the reduced chi-squared of
the column-density residuals,
\begin{equation}
  \chi^2_\nu = \frac{1}{n-k}\sum_i
  \frac{\big[\log N_i - \log N(b_i)\big]^2}{\sigma_{\log N_i}^2},
  \label{eq:chi2}
\end{equation}
where $n$ is the number of selected spaxels ($n = 69$ for the fiducial
S/N$\,>2$ selection), $k= 2$, and $\log N(b_i)$ is the (fixed) noiseless
best-fit model. Since the input model contains no intrinsic dispersion, a value
$\chi^2_\nu \approx 1$ indicates that the residuals are consistent with the injected
uncertainties, whereas $\chi^2_\nu > 1$ signals additional, unaccounted-for
scatter. We quote the median $\chi^2_\nu$ and corresponding dispersion over the
$3\,000$ realisations. We caution that the flux-contribution maps of neighbouring spaxels overlap
(Sect.~\ref{sec:formalism}), so the data points are not strictly independent and
the effective number of degrees of freedom is somewhat smaller than $n-k$. The
absolute $\chi^2_\nu$ values are therefore only indicative; because the same
convention is applied to both coordinate definitions, the comparison between them
is unaffected.

\subsection{PSF misspecification}
\label{app:psf-test}
 
The flux-contribution maps $w_a$ depend on the assumed PSF. To quantify the effect of
an imperfect PSF model, we recomputed the flux-weighted coordinates with the kernel
FWHM misspecified by $\pm10\%$ ($\theta_{\rm PSF} = 0\farcs63$ and $0\farcs77$)
about the
nominal $\theta_\text{PSF}=0\farcs7$ adopted in Sect.~\ref{sec:simulation}, rebuilding the
matching convolution kernel and
repeating the same fitting procedure described above. The spaxel-centre coordinates are purely 
geometric and are therefore identical across experiments. The results are summarised in
Table~\ref{tab:sp_fw:noisy}. A $\pm10\%$ error in the PSF FWHM shifts the recovered
$b_e$ by $<1\%$ and $\log N_0$ by $\lesssim 0.02$~dex and leaves $\chi^2_\nu \approx 1$ 
in every case. The flux-weighting is thus robust to realistic PSF misspecification. 

\subsection{Spaxel selection}
\label{app:selection-test}
 
\begin{figure}[h]
    \centering
    \includegraphics[width=\linewidth]{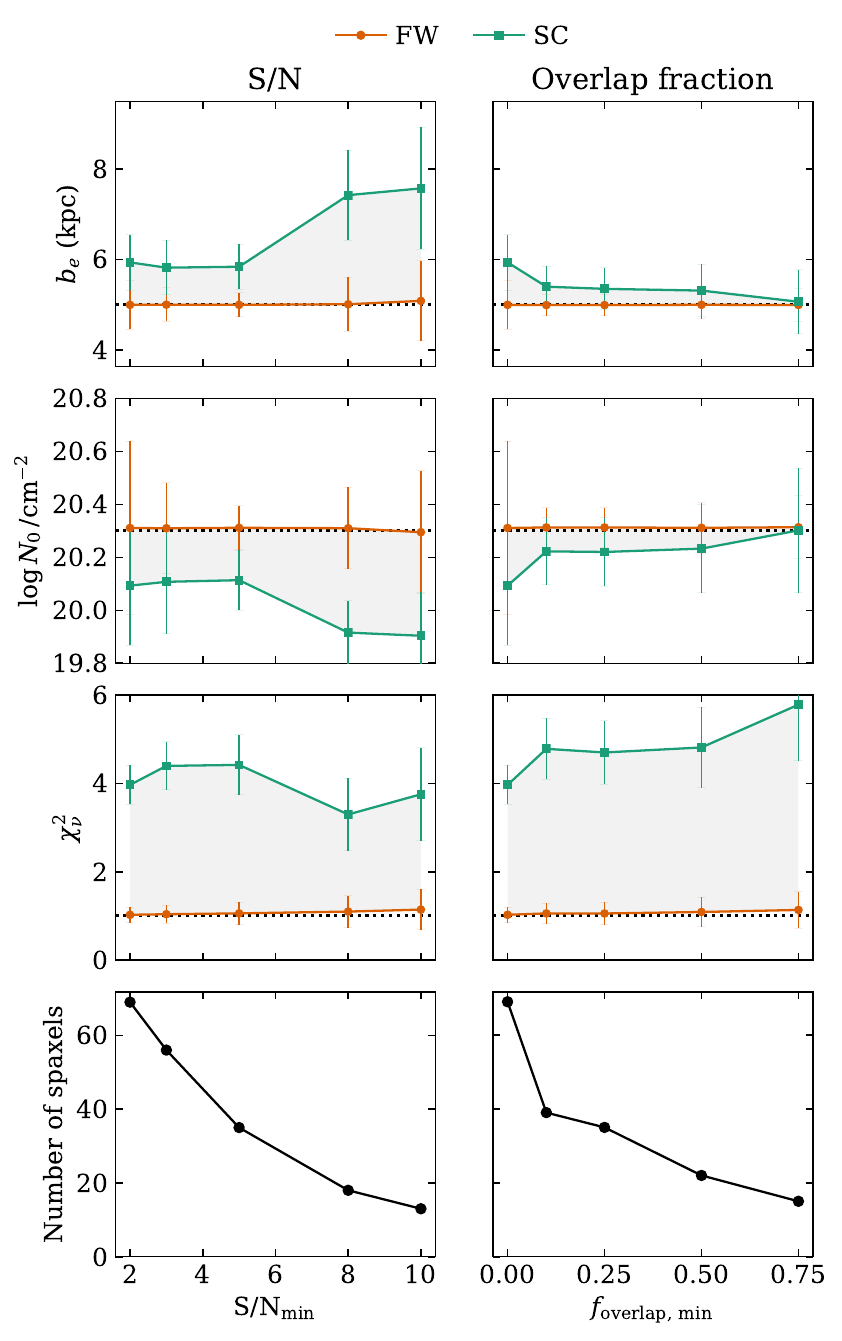}
    \caption{
    Recovered $b_e$, $\log N_0$, $\chi^2_\nu$, and number of selected spaxels as a function of two spaxel-selection thresholds.
    \textit{Left panels:} S/N thresholds.
    \textit{Right panels:} Minimum fraction of the spaxel
    area
    overlapping the unconvolved arc continuum footprint 
    ($f_{\mathrm{overlap},\min}$).
    The orange and green lines represent the FW and SC datasets, respectively.
    The bottom row displays the number of selected spaxels 
    for each threshold. The horizontal dotted lines mark 
    the input model values ($b_e = 5.0$~kpc, 
    $\log N_0/\text{cm}^{-2} = 20.3$) for the first two rows and $\chi^2_\nu=1$ for the third row.
}
    \label{fig:fits-results-spaxel-selection}
\end{figure}

Arc-tomography analyses conventionally select spaxels with an S/N threshold. We
found, however, that the displacement vectors are larger for off-arc spaxels ($f\sim0$). 
We therefore explored two criteria: a
minimum S/N ($2, 3, 5, 8, 10$) and a minimum $f$ ($0, 0.1, 0.25, 0.5, 0.75$).
Figure~\ref{fig:fits-results-spaxel-selection} shows $b_e$, $N_0$, 
$\chi^2_\nu$, and the number of selected spaxels as a function
of each threshold, for the noise-injected column densities. 

Imposing any non-zero $f$ removes spaxels that lie entirely 
outside the
arc footprint. For the spaxel-centre coordinates, this pulls the
recovered parameters
towards the input values ($b_e:\, 5.9 \to 5.1$~kpc 
and $\log N_0/\mathrm{cm^{-2}}:\, 20.08 \to 20.30$ as the threshold increases), 
bringing them into $\sim\!1\sigma$ agreement. However, increasing
the cut in $f$ does not improve $\chi^2_\nu$, and it 
remains $\approx 4$--$6$. The flux-weighted fit, on the other hand, 
is insensitive to the cut:
$b_e$ and $N_0$ remain at the input values with
$\chi^2_\nu \approx 1$ throughout.
 
An S/N cut behaves very differently for the spaxel-centre coordinates. 
Because integrated flux is not strictly determined by whether a 
spaxel overlaps with the arc, raising the threshold does not guarantee
removing the most misassigned spaxels, and, at high S/N, actively
worsens the bias: $b_e$ rises to $\sim\!7.5$~kpc and $N_0$ falls to
$\sim\!0.8\times10^{20}\,\mathrm{cm^{-2}}$ at S/N$\,>10$, 
with $\chi^2_\nu$ remaining $\approx 3$--$4$. In contrast, the flux-weighted
fit recovers the input parameters with $\chi^2_\nu \approx 1$ at every 
threshold.

\subsection{Spaxel size}
\label{app:binning-test}

\begin{table}[b]
\footnotesize
\centering
\setlength{\tabcolsep}{4pt}
\caption{Recovery of the input parameters ($b_e\!=\!5.0$~kpc;
$\log N_0/\mathrm{cm}^{-2}\!=\!20.3$) as a function of the spatial binning, at
fixed $\theta_\text{PSF}=0\farcs7$, for the SC and FW datasets, with no mock uncertainties in $\log \bar{N}$. }
\label{tab:binning}
\begin{tabular}{lcccc}
\toprule
 & \multicolumn{2}{c}{SC} & \multicolumn{2}{c}{FW} \\
\cmidrule(lr){2-3}\cmidrule(lr){4-5}
Bin & $b_e$ & $\log N_0/\mathrm{cm^{-2}}$ & $b_e$ & $\log N_0/\mathrm{cm^{-2}}$ \\
(\arcsec) & (kpc) & & (kpc) & \\
\midrule
$0.2$ & $5.7 \pm 0.1$ & $20.13 \pm 0.02$ & $4.99 \pm 0.04$ & $20.31\pm0.01$ \\
$0.4$ & $5.5\pm0.1$ & $20.19\pm0.02$ & $4.98\pm0.06$ & $20.31\pm0.02$ \\
$0.6$ & $5.5\pm0.2$ & $20.20\pm0.04$ & $4.98\pm0.10$ & $20.31\pm0.03$ \\
$0.8$ & $5.7 \pm 0.3$ & $20.14 \pm 0.05$ & $4.99 \pm 0.07$ & $20.31 \pm 0.02$ \\
\bottomrule
\end{tabular}
\end{table}

The fiducial $0\farcs8$ binning of Sect.~\ref{sec:simulation} follows
\tejos{} in matching the spaxel size to $\theta_\text{PSF}$. A natural
question is whether finer sampling, at the same seeing, improves the profile recovery.
We therefore repeated the full experiment at $0\farcs6$,
$0\farcs4$, and $0\farcs2$ (the native MUSE spaxel scale), rebuilding $L$,
the masks $M_a$, the sensitivity maps $S_a$, and the weights $w_a$ at each
scale, while holding $\theta_\text{PSF}$, the arc segmentation mask, the
lens model, and the input column-density profile fixed. 
The spaxel selection ($>\!50\%$ of the integrated flux
from the arc and S/N$\,>2$) was applied unchanged at each scale. We then
recalculated and fitted the corresponding noiseless $\log \bar{N}$ values. 

The results are given in Table~\ref{tab:binning}. The bias does not
diminish with finer sampling: the SC fit continues to overestimate $b_e$ 
and underestimate $\log N_0$, while the FW fit
recovers the input parameters throughout. The reason is contained in
Eq.~\eqref{eq:sensitivity-map}. As the spaxel shrinks below the PSF, $M_a$
tends to a delta function and $S_a \rightarrow K$, so the
region a spaxel actually samples saturates at the PSF width, independently 
of the bin size. Finer binning also carries two costs: per-spaxel uncertainties
grow as the per-spaxel flux diminishes, and neighbouring spaxels become
increasingly correlated. Both effects argue for retaining a
PSF-matched binning and correcting the coordinates, rather than sampling
more finely in the hope of avoiding the bias.

\end{appendix}

\end{document}